# An ab initio non–equilibrium Green's function approach to charge transport: dithiolethine

Alexander Schnurpfeil[a,b], Bo Song[c*], and Martin Albrecht[a]

[a] *Theoretical Chemistry FB08 - University Siegen - 57068 Siegen/Germany,*

[b] *Institute for Theoretical Chemistry - University of Köln - 50939 Köln/Germany,*

[c] *Department of Physics, Beijing Technology and Business University, Beijing 100037*

## Abstract

We present a novel ab initio non–equilibrium approach to calculate the current across a molecular junction. The method rests on a wave function based full ab initio description of the central region of the junction combined with a tight binding approximation for the electrodes in the frame of the Keldysh Green's function formalism. Our procedure is demonstrated for a dithiolethine molecule between silver electrodes. The main conducting channel is identified and the full current–voltage characteristic is calculated.

---

[*] Corresponding author: e-mail: spaul@netease.com, Phone: +86-10 13521001463

Recent years have seen a steep rise in the broad field of nano engineering with molecular junctions being a significant part of it[1]. This has been brought about by tremendous advances in engineering techniques, which led to both unraveled reduction in size and variety[2]. The particular interest in molecular junctions is to ultimately design switches or 'transistors' on a nano scale which might be triggered by phenomena other than electrical current.

In outstanding experiments a current through a single molecule has been observed employing either a gold tip[3] as well as indeed fixing a molecule chemically well defined between two gold electrodes[4]. However, at present experiments are still notoriously difficult and hard to interpret, thus shifting weight to theoretical considerations which are expected to both create a fundamental understanding of the microscopic processes involved and provide a guidance for improved engineering. As miniaturization reached an extend which invokes the principles of quantum mechanics, quantum chemical methods are now being applied to nano structures so as to boost our understanding of small heterogeneous devices.

Method development and applications geared towards a genuine description of transport phenomena across molecular junctions has just started to form a new and promising branch of modern quantum chemistry and theoretical physics. Theoretical descriptions of the problem try to illuminate partial aspects like the role of the molecular electronic structure[5–7] or the influence of various structural conformations[8–10].

Some progress has been made recently based on the local density approximation (LDA) to density functional theory (DFT) as a starting point, which provides numerically affordable applications to the molecular junction problem[5–11]. Further approximations are commonly built on top of LDA, like the tight binding approach (TB) or parametrized minimal basis sets[12]. Another set of approaches renounces completely attempts of *ab initio* calculations and resorts to empirical models[13–16].

A more advanced scheme developed by Xue, Ratner and Datta sticks with these approximations, but develops a non–equilibrium formalism[14,16–18]. The earlier attempts were presented by Wang, Guo and Taylor.[19–21]. The approaches above are almost based on model Hamiltonian technique or DFT.

Wave function based *ab initio* methods typically display a steeper increase in numerical cost with system size, but offer a straightforward and systematic applicability, with eigenvalues being obtainable, in principle, to any desired accuracy. Recently one of the authors developed a completely wave function based *ab initio* procedure to obtain the Green's function and related quantities for solids, polymers and molecules[22–25]. Subsequently this scheme proved valiant for the calculation of the transmission coefficient in the frame of the zero–voltage approximation to the Landauer Theory[26,27]. While these efforts where limited to the equilibrium Green's function, one of the authors recently presented a model implementation of the Keldysh non–equilibrium theory and obtained the full current–voltage characteristic for a quantum dot model[28]. The procedure relies on a formulation of Haug and Jauho[29] and replaces the equilibrium Green's function by the Keldysh Green's function which is obtained from a time integration in the complex time plane rather than the real time axis, thus accounting for non–equilibrium effects brought about by a coupling of the system under consideration to a reservoir. A derivation is given by Rammer and Smith[30].

The idea of the present work is to combine the accuracy of a full fledged wave function based *ab initio* method with the Keldysh formalism as it was established for charge transport through molecular junctions. To this end we stick to the tight binding approximation of Ref.[28] as far as the electrodes are concerned, but replace the quantum dot by a realistic organic molecule which is treated together with its contacts to the electrodes on a full quantum chemical basis.

A sketch of the system we investigated is given in Fig. 1. A dithiolethine molecule is put between two silver electrodes so that the thiol bridges bind covalently to two silver atoms, which replace the hydrogens originally present in dithiolethine. The silver electrodes (grey–shaded in the figure) are modeled as semi–infinite one–dimensional chains in the frame of the tight binding approximation. The central region (dashed box in the figure) is treated in a fully *ab initio* way. It contains the dithiolethine molecule as well as one silver atom on each side.

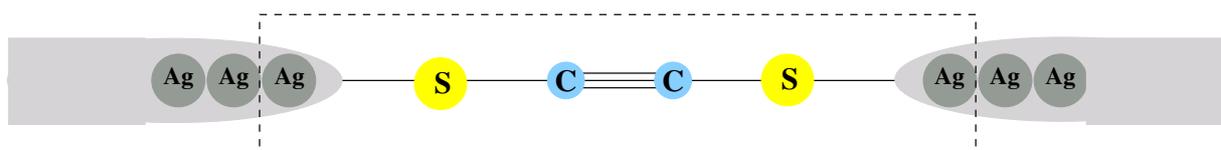

FIG. 1: *Sketch of the dithiolethine between two silver electrodes modeled as one–dimensional chains (shaded regions). The first silver atom to the left and right of the molecule replaces the hydrogens of dithiolethine and forms part of the central region (dashed box).*

The central region is treated in a fully *ab initio* way. First the structure of the bare dithiolethine molecule is obtained by geometry optimization on the B3LYP level. Subsequently the system of the central region is constructed by replacing the hydrogen atoms of the original dithiolethine molecule by the silver atoms to the left and right with the Ag–S distance taken to be the average chemical value of 2.74 Å. For this system canonical Hartree–Fock (HF) orbitals are calculated employing a STO–3G basis with the program package MOLPRO[31]. These orbitals are subsequently localized by means of the Pipek–Mezey option. By virtue of the program package MOLCAS a complete one– and two–electron integral transformation is performed in these orbitals, resulting in the Fock matrix F and the standard two–electron integrals W. Finally the occupied and virtual HF orbitals $\chi$ relating to the silver electrodes are projected out from the remaining occupied as well as virtual molecular HF orbitals which are denoted by indices m and n henceforth. All orbital indices are understood to also carry the spin index throughout this work. The coupling matrices $H_{m\chi}^{L}$ and $H_{m\chi}^{R}$ are computed from these orbitals, where we adopt the convention that L and R refer to the left and right electrode, respectively. Electronic correlations are then included by means of the program package GREENS developed by one of the authors[22–27]. To this end the self energy is constructed in a fully *ab initio* way.

In terms of the local HF orbitals a model space P and excitation space Q are distinguished for the example of virtual states (the case of occupied states being completely analogous) as follows: The model space P describing the HF level comprises of the $(N+1)$–particle HF determinants $|n\rangle$, while the correlation space Q contains single (and , in principle double) excitations $|\beta\rangle$ on top of $|n\rangle$:

$$|n\rangle = c_n^\dagger |\Phi_{\mathrm{HF}}\rangle, \qquad |\beta\rangle = c_r^\dagger c_a |n\rangle, \quad c_r^\dagger c_s^\dagger c_a c_b |n\rangle \tag{1}$$

$$P = \sum_n |n\rangle\langle n|, \qquad Q = \sum_\beta |\beta\rangle\langle \beta|. \tag{2}$$

We adopt the index convention that $a, b, c, d$ and $r, s, t, u$ represent occupied and virtual HF orbitals, respectively.

The Green's function $G_{\mathrm{nm}}(t) = -i\langle T[c_n(0)c_m^\dagger(t)]\rangle$, where T is the time–ordering operator and the brackets denote the average over the exact ground–state, can be obtained from Dyson's equation as:

$$G_{nm}(\epsilon) = [\epsilon \underline{1} - F - \Sigma(\epsilon)]_{\mathrm{nm}}^{-1}. \tag{3}$$

Here the self energy $\Sigma_{\mathrm{nm}}(\epsilon)$ which contains the correlation effects, has been introduced and $\underline{1}$ represents the unity matrix. To construct the self energy the resolvent

$$[\epsilon \underline{1} - H^r + i\delta \underline{1}]_{\beta;\beta'}^{-1} \tag{4}$$

is needed. It can be gained from diagonalization of the Hamiltonian

$$[H^{\mathrm{r}}]_{\beta,\beta'} = \langle\beta|H - E_0|\beta'\rangle, \tag{5}$$

where the states $|\beta\rangle, |\beta'\rangle$ are those of the correlation space Q as in Eq. (1). Here $E_0$ is the HF ground state energy while the brackets indicate the HF average.

The self energy is approximated by decomposition into a retarded and an advanced part. In what follows the superscript $^{\mathrm{r}}$ will be used throughout to refer to the retarded case, while $^{\mathrm{a}}$ is taken to denote the advanced part. Furthermore, the configuration space will be restricted to single excitations, i. e. three–body–interactions.

In the following only the construction of the retarded self energy part is given, the case of the advanced part being analogous.

The space of 2–particle 1–hole states (2p1h) is spanned by $|r, s, a\rangle = a_r^\dagger a_s^\dagger a_a |\Phi_{\mathrm{HF}}\rangle$. The Hamiltonian is set up in this basis as: $[H^{\mathrm{r}}]_{rsa,r's'a'} = \langle r, s, a|H - E_0|r', s', a'\rangle$ and is subsequently diagonalized.

Diagonalizing the matrix $H^{\mathrm{r}}$ results in the eigenvectors $S^{\mathrm{r}}$ and eigenvalues $\lambda^{\mathrm{r}}$. The retarded part of the self energy is then constructed as

$$\begin{aligned}\Sigma_{\mathrm{nm}}^{\mathrm{r}}(\epsilon) &= \sum_{\mathrm{rsa;r's'a'}} \Upsilon(\mathrm{rs;na})[\epsilon\underline{1} - H^{\mathrm{r}} + i\delta\underline{1}]^{-1}_{\mathrm{rsa;r's'a'}} \Upsilon(\mathrm{r's';ma'}) \\ &= \sum_{\mathrm{rsa;r's'a'}} \Upsilon(\mathrm{rs;na}) \sum_{\mathrm{q}} S^{\mathrm{r}}_{rsa;q} \frac{1}{(\epsilon - \lambda^{\mathrm{r}}_{\mathrm{q}} + i\delta)} S^{\mathrm{r}}_{q;r's'a'} \Upsilon(\mathrm{r's';ma'}).\end{aligned} \tag{6}$$

$\Upsilon$ is a shorthand for $\Upsilon(\mathrm{rs;ta}) = W_{\mathrm{rsta}} - W_{\mathrm{rsat}}$ and $W$ are the standard two–electron integrals.

This procedure is now combined with the formalism presented for a one–level quantum dot model by Yang et al.[28]. The current is given according to Ref.[28] by

$$J = \frac{ie}{\hbar} \int \frac{d\epsilon}{2\pi} [f_L(\epsilon) - f_R(\epsilon)] \cdot \Gamma_L(\epsilon) \left[\Gamma_L(\epsilon) + \Gamma_R(\epsilon)\right]^{-1} \Gamma_R(\epsilon) \cdot [G^r(\epsilon) - G^a(\epsilon)], \tag{7}$$

where the linewidth functions to the left or right ($\alpha =$L or $\alpha =$R) have the form

$$\Gamma_{\alpha;\mathrm{mn}}(\epsilon) = H^\alpha_{\mathrm{m}\chi} H^\alpha_{\chi\mathrm{n}} 2\pi \sum_k \cos^2(k) \cdot \delta(\epsilon - \epsilon_{k,\alpha}). \tag{8}$$

The overall linewidth function is just the sum:

$$\Gamma_{\mathrm{mn}} = \Gamma_{L;\mathrm{mn}} + \Gamma_{R;\mathrm{mn}}. \tag{9}$$

The energy eigenvalues $\epsilon_{k,\alpha}$ of the electrodes are given according to the tight binding approximation by

$$\epsilon_{k,\alpha} = -2t\cos(k) + eV_\alpha, \tag{10}$$

where $V_\alpha$ is the external voltage applied to electrode $\alpha$. In fact we take $V_L = V$ and $V_R = 0$, where V is the external potential applied across the junction. The tight binding parameter t is taken to be half the band width of a silver chain which was reported by Springborg and Sarkar to be 5 eV[32].

The Fermi functions $f_\alpha(\epsilon)$ are defined as in Ref.[28]:

$$f_\alpha(\epsilon) = \Theta(\epsilon - D - eV_{\text{alpha}})\Theta(\epsilon - \mu_F - eV_{\text{alpha}}). \tag{11}$$

Here $D = -2t$ and $\mu_F = 0$ are taken to be the bottom of the energy band and the Fermi energy in the leads.

Finally the Green's function including electron correlations and the linewidth function is calculated from:

$$G^r_{mn}(\epsilon) = [\epsilon - F_{mn} - \Sigma_{mn}(\epsilon) + i\Gamma_{mn}(\epsilon)/2]^{-1}. \tag{12}$$

The extension of the quantum dot approach towards this *ab initio* treatment is visible by the replacement of the simple dot–electrode coupling parameter t' in Ref.[28] by the coupling matrices $H^L_{m\chi}$ and $H^R_{m\chi}$ in Eq. (8) as well as the inclusion of the full correlation treatment for the dithiolethine by means of the self energy matrix $\Sigma_{mn}(\epsilon)$ appearing in Eq. (12).

For numerical evaluation of the current in Eq. (7) all matrices are transformed to the eigenstates of the linewidth matrix $\Gamma$, which are identified with the eigenchannels of the system with respect to conduction. Eigenvalues of less than 1 $\mu$Hartree were ignored.

Our result for the current–voltage characteristic is shown in Fig. 2. The overall curve displays the typical characteristics found earlier for the one–level quantum dot model[28], featuring essentially one peak. This is consistent with the fact that upon diagonalizing the linewidth matrix $\Gamma$ we only found one eigenchannel. Moreover we found that the corresponding eigenvector only has one major component at a virtual level of dithiolethine with eigenenergy 5.89 eV. This position is marked by an arrow in Fig. 2. Consequently we observe a steep increase of the current–voltage characteristic at this point. The drop–off at higher voltage finally is forced by the finite width of the silver bands in the electrodes. In fact it was established in Ref.[28] that the finite band width effect sets in at about 4t, in our case this amounts to 10 eV, and indeed a sharp drop of the current is observed at this point in the figure. Quantitatively we note that the order of magnitude of 0.1 $\mu$A is in the range to be expected on the basis of experimental data on related systems. Measurements on anthracene derivates suggested similar values for the current[33], which was found to increase up to 0.5 $\mu$A when gold electrodes where employed.

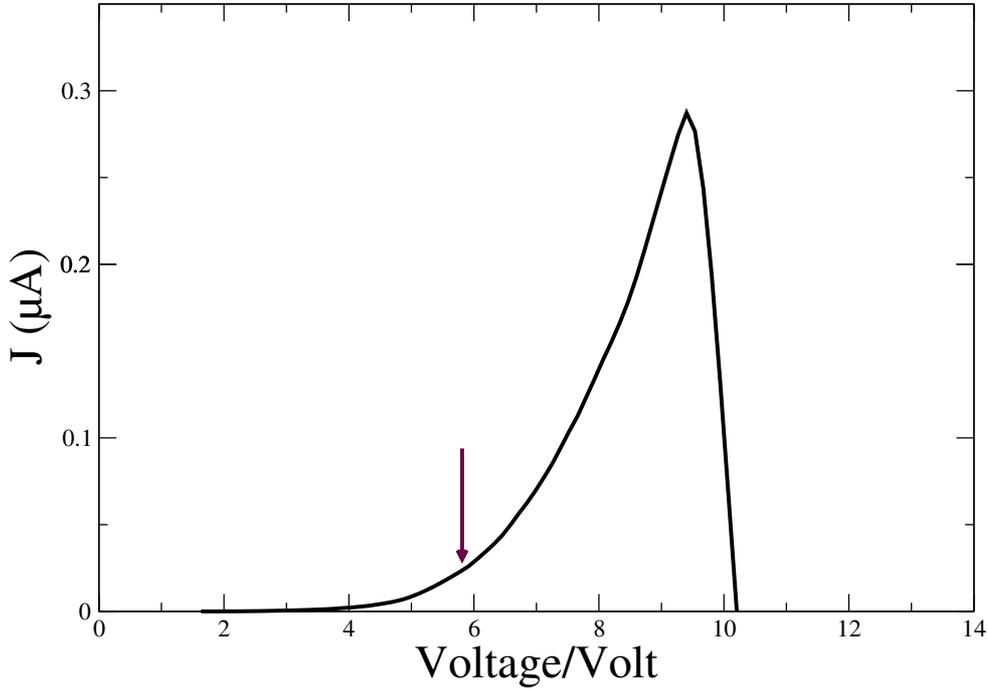

FIG. 2: *Current–voltage characteristic for dithiolethine between silver electrodes as sketched in Fig. 1. The arrow marks the position at which the main component of the conduction channel is located energetically.*

In sum we presented a novel method to compute the full current–voltage characteristic of a molecular junction. While the electrodes are still treated on the tight–binding level, a full–scale wave function based *ab initio* calculation is performed for the central region. Both the coupling of the molecular states to the electrodes as well as electronic correlations on the molecular system have been taken into account. The results enter the current both via the Green's function and the linewidth functions in the frame of the non–equilibrium Keldysh formalism. The results are transparent and consistent with earlier findings both qualitatively and quantitatively.

We believe that this is a breakthrough which will lead to large scale applications in the future.

## I. ACKNOWLEDGMENTS

The authors are grateful for support from the German Research Foundation (DFG) in the frame of the programs SPP 1145 and AL625/2–1.

---